\documentclass[preprint,showpacs,preprintnumbers,amsmath,amssymb,floatfix,endfloats*]{revtex4}
\usepackage{graphicx}% Include figure files
\usepackage{dcolumn}% Align table columns on decimal point
\usepackage{bm}% bold math
\usepackage{amsfonts}

\DeclareMathAlphabet{\mathsfsl}{OT1}{cmr}{bx}{it}
\begin{document}
%----------------------------------------------------------------------%
% Title
%----------------------------------------------------------------------%
\title{Plastic deformation of a model glass induced by a local shear transformation}
\author{Nikolai V. Priezjev}
\affiliation{Department of Mechanical and Materials Engineering,
Wright State University, Dayton, OH 45435}
\date{\today}
\begin{abstract}

The effect of a local shear transformation on plastic deformation of
a three-dimensional amorphous solid is studied using molecular
dynamics simulations.    We consider a spherical inclusion, which is
gradually transformed into an ellipsoid of the same volume and then
converted back into the sphere.     It is shown that at sufficiently
large strain amplitudes, the deformation of the material involves
localized plastic events that were identified based on the relative
displacement of atoms before and after the shear transformation. We
found that the density profiles of cage jumps decay away from the
inclusion, which correlates well with the radial dependence of the
local deformation of the material.     At the same strain amplitude,
the plastic deformation becomes more pronounced in the cases of
weakly damped dynamics or large time scales of the shear
transformation.  We showed that the density profiles can be
characterized by the universal function of the radial distance
multiplied by a dimensionless factor that depends on the friction
coefficient and the time scale of the shear event.

\end{abstract}

\pacs{62.20.F-, 61.43.Fs, 83.10.Rs}

% \pacs{61.43.Fs, 61.43.-j, 62.20.F-, 83.10.Rs}

%   64.70.pj Polymers
%   81.05.Kf, Glasses (including metallic glasses)
%   61.43.Fs Glasses
%   64.70.P-  Time-dependent properties; relaxation (for glass transitions)
%   61.43.-j  Disordered solids, structure
%   66.30.Pa  Diffusion in nanoscale solids
%   62.20.F-  Deformation: mechanical properties of solids
%   83.10.Rs  Computer simulation of molecular and particle dynamics

\maketitle

\section{Introduction}

Despite its long history, the prediction of mechanical response of
amorphous materials, such as metallic or polymer glasses, remains an
area of active research~\cite{FalkRev10,ChengMa11}.      The main
challenge in this field is to fully characterize the dynamics of
atomic scale rearrangements during plastic deformation of
glasses~\cite{BarratCh8}. These localized rearrangements were first
observed by Argon and Kuo in two-dimensional bubble
rafts~\cite{Argon79,ArgonKuo}. Later, Falk and Langer proposed a
dynamical mean-field theory of low temperature shear deformation in
amorphous solids in terms of the shear transformation zones, which
accounts for many of the features seen in simulations, including
strain hardening and yield stress~\cite{Falk98}.     A number of
studies considered various criteria in order to identify regions
susceptible to plastic rearrangement by examining the local
density~\cite{Bailey06}, elastic
moduli~\cite{dePablo04,Tsamados09,Procaccia10},  short range
order~\cite{Teichler02,Albano05}, and ``soft spots" from the
low-frequency vibrational
modes~\cite{Tanguy10,Manning11,Schoenholz14}.    Athermal
quasistatic simulations have clearly shown that heterogeneous
plastic flow of two-dimensional amorphous solids involves
quadrupolar localized rearrangements and system spanning shear
bands~\cite{Maloney06,Tanguy06,Tsamados10,Maloney12}.

\vskip 0.05in

Twenty years ago, Argon and Bulatov~\cite{Bulatov94} developed a
model where plastic flow was treated as a stochastic sequence of
local inelastic rearrangements that interact via a long-range
elastic field~\cite{Eshelby57}.  In this model, each local
rearrangement results in a global stress drop and stress
redistribution in its neighborhood, which in turn may drive nearby
regions towards their instability threshold and cause secondary
plastic events~\cite{Bulatov94}.    In particular, a transition from
diffuse plastic flow at high temperatures to localized flow through
shear bands at low temperatures was observed~\cite{Bulatov94}.
Recent molecular dynamics simulations of a two-dimensional amorphous
system have shown that at finite shear rates long-range interactions
between local rearrangements give rise to directional quasilinear
avalanches~\cite{Lemaitre09}.

\vskip 0.05in

More recently, the elastic response of a two-dimensional amorphous
solid to a local shear transformation was studied using molecular
dynamics simulations~\cite{Puosi14}.  The fictitious transformation
due to an instantaneous displacement of about twenty particles
within a circular region replicated an elementary plastic event in a
sheared system.   It was found that at small values of the shear
strain, the averaged displacement field in the material agrees well
with the predictions of the continuum elasticity theory in both the
stationary and transient regimes~\cite{Puosi14}.     It was also
shown that the propagation of the elastic signal varies from a
diffusive transmission for strong damping to a propagative
transmission in the case of weakly damped dynamics.   One of the
motivations of the current study is to investigate the effect of a
local shear transformation on plastic deformation of a
three-dimensional model glass.

\vskip 0.05in

The mechanical response of amorphous materials to oscillatory shear
deformations was recently investigated
experimentally~\cite{Arratia13,Arratia14,Spaepen14,Cipelletti14} and
by means of atomistic
simulations~\cite{Sastry13,Priezjev13,Reichhardt13,Priezjev14}. It
was found that at small strain amplitudes, the amorphous systems
gradually settle into dissipative limit cycles where particles are
displaced periodically but then return to their original
positions~\cite{Sastry13,Reichhardt13,Arratia14}.  Interestingly,
the deformation process involves a number of localized plastic
events that are reversible with each
cycle~\cite{Arratia13,Reichhardt13,Arratia14}.   With increasing
strain amplitude above a critical value, the rearrangement of
particles becomes irreversible, leading to diffusive
behavior~\cite{Sastry13,Priezjev13,Reichhardt13,Arratia14,Priezjev14}.
It was also shown that the structural relaxation process at finite
temperatures involves intermittent bursts of clusters of particles
undergoing large
displacements~\cite{Priezjev13,Cipelletti14,Priezjev14}.

\vskip 0.05in

In this paper, molecular dynamics simulations are carried out to
study structural relaxation in an amorphous solid induced by a local
shear transformation.   We introduce a reversible transformation of
a spherical inclusion in a quiescent system.   From a physical point
of view, this situation corresponds to a vibrating inclusion within
an amorphous solid (for example, a cyclic thermal or mechanical
deformation of an embedded inclusion in an amorphous matrix).   The
cage detection algorithm is used to identify large particle
displacements during the shear event.    The spatial distribution of
cage jumps is analyzed for different damping conditions, strain
amplitudes, and duration of the shear event, and then correlated
with the amplitude of the local deformation of the material.

\vskip 0.05in

The rest of the paper is organized as follows.     The description
of molecular dynamics simulation model is given in the next section.
In Sec.\,\ref{sec:Results}, the results for the radial density
profiles of cage jumps as a function of the strain amplitude, the
shear transformation time scale, and the friction coefficient are
presented. The conclusions are provided in the final section.

\section{Details of molecular dynamics simulations}
\label{sec:MD_Model}

\begin{figure}[t]
\includegraphics[width=14.cm,angle=0]{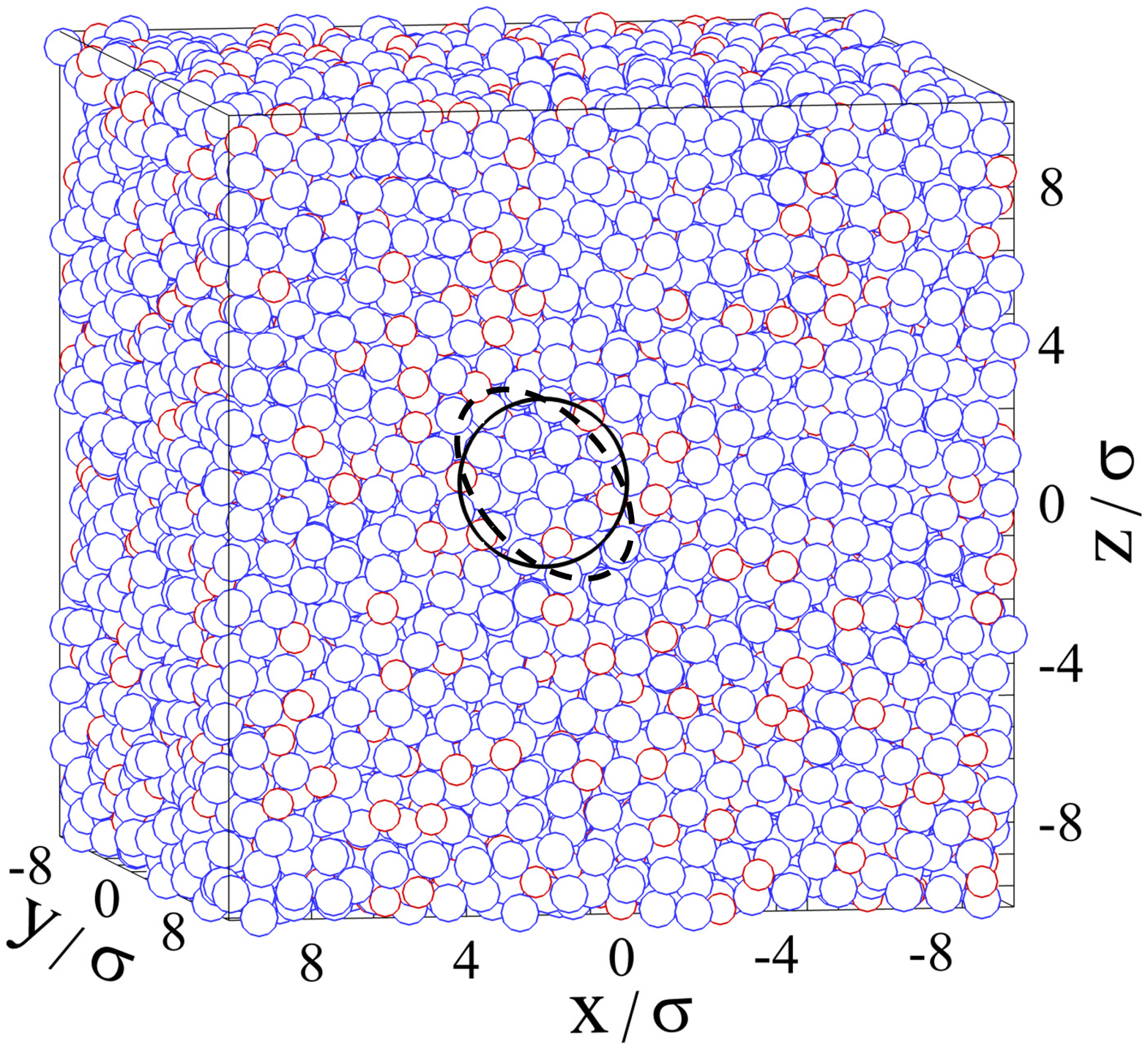}
\caption{(Color online)   A snapshot of the equilibrated binary LJ
glass in a periodic box.   Atoms of type A are indicated by large
blue circles and atoms of type B are denoted by small red circles.
Atoms within a sphere (black circle) form an inclusion.    The local
shear transformation is introduced by displacing the inclusion atoms
into an ellipsoid of the same volume (dashed ellipse). }
\label{fig:snapshot_system}
\end{figure}

We consider a three-dimensional system composed of $N=10\,000$
particles in a periodic box (see Fig.\,\ref{fig:snapshot_system}).
The model glass is represented by the Kob-Andersen (KA) binary
(80:20) mixture with non-additive interaction parameters of the
Lennard-Jones (LJ) potential in order to avoid
crystallization~\cite{KobAnd95}.   In the KA model, particles
$\alpha,\beta=A,B$ interact via the pairwise LJ potential
\begin{equation}
V_{\alpha\beta}(r)=4\,\varepsilon_{\alpha\beta}\,\Big[\Big(\frac{\sigma_{\alpha\beta}}{r}\Big)^{12}\!-
\Big(\frac{\sigma_{\alpha\beta}}{r}\Big)^{6}\,\Big],
\label{Eq:LJ_KA}
\end{equation}
with the parameters $\varepsilon_{AA}=1.0$, $\varepsilon_{AB}=1.5$,
$\varepsilon_{BB}=0.5$, $\sigma_{AB}=0.8$, $\sigma_{BB}=0.88$, and
$m_{A}=m_{B}$.   In all simulations, the cutoff radius was chosen to
be twice the minimum position of the LJ potential, i.e.,
$r_{c,\,\alpha\beta}=2.245\,\sigma_{\alpha\beta}$.   The units of
length, mass, and energy are defined $\sigma=\sigma_{AA}$,
$m=m_{A}$, and $\varepsilon=\varepsilon_{AA}$, and, therefore, the
unit of time is $\tau=\sigma\sqrt{m/\varepsilon}$.

\vskip 0.05in

The time evolution of the system is described by the Langevin
dynamics. For example, the equation of motion in the $\hat{x}$
direction is given by
\begin{equation}
\label{Eq:Langevin_x} m\ddot{x}_i + m\Gamma\dot{x}_i = -\sum_{i \neq
j} \frac{\partial V_{ij}}{\partial x_i} + f_i\,,
\end{equation}
where $V_{ij}$ is the total interaction potential, $\Gamma$ is the
friction coefficient, and $f_i$ is a random force with zero mean and
variance $\langle
f_i(0)f_j(t)\rangle=2mk_BT\Gamma\delta(t)\delta_{ij}$ determined by
the fluctuation-dissipation theorem.   The Langevin thermostat
temperature is set $T=10^{-2}\,\varepsilon/k_B$, where $k_B$ is the
Boltzmann constant. The equations of motion were integrated using
the fifth-order Gear predictor-corrector algorithm~\cite{Allen87}
with a time step $\triangle t_{MD}=0.005\,\tau$.

\vskip 0.05in

The model glass was confined into a cubic box with a fixed side
length $L=20.27\,\sigma$ so that the total density
$\rho=\rho_{A}+\rho_{B}=1.2\,\sigma^{-3}$ remained constant.   The
glass transition temperature for the KA binary mixture at this
density is about $0.45\,\varepsilon/k_B$~\cite{KobAnd95}.   In our
setup, periodic boundary conditions were implemented along all three
directions.    At first, the system was equilibrated for about
$5\times10^6$ MD steps at the temperature $1.1\,\varepsilon/k_B$ and
then gradually quenched with the rate $10^{-5}\,\varepsilon/k_B\tau$
to the final temperature $10^{-2}\,\varepsilon/k_B$.    The
post-processing analysis was performed in $504$ independent samples.

\vskip 0.05in

% deformation protocol

The plastic deformation of the material was induced by a fictitious
local shear transformation.   In our simulations, the inclusion
consists of about 135 atoms within a sphere of radius
$r_i=3\,\sigma$, which is located at the center of the simulation
box (see Fig.\,\ref{fig:snapshot_system}).    The system was first
aged for about $500\,\tau$ at the temperature
$10^{-2}\,\varepsilon/k_B$, while the atoms within the inclusion
were kept fixed.     Then, the inclusion atoms were gradually
displaced so that a sphere was transformed into an ellipsoid of the
same volume.    During this procedure, the major axis of the
ellipsoid was always oriented parallel to the $(1,1,1)$ direction,
as shown in Fig.\,\ref{fig:snapshot_system}.   In the following, the
ratio of the length of the ellipsoid semi-major axis and the sphere
radius $r_i$ defines the shear strain of the transformation.  The
shear strain was varied during the time interval $0\leqslant t
\leqslant \tau_i$ according to the following equation
\begin{equation}
\epsilon\,(t) = \epsilon_0\,\,\textrm{sin}(\pi t/\tau_i),
\label{Eq:strain}
\end{equation}
where $\epsilon_0$ is the strain amplitude and $\tau_i$ is the time
scale of the shear transformation.      After $t=\tau_i$, the
inclusion atoms were fixed again at their original positions (before
the transformation) and the system was equilibrated for additional
$10^3\,\tau$, and then the average positions of all atoms were
computed again.     For each independent sample, the averaged atom
positions before and after the shear transformation were stored and
then analyzed to determine plastic deformation of the material.

\section{Results}
\label{sec:Results}

% general strain

In the absence of mechanical deformation, the atomic structure of
the model glass lacks the long-range order characteristic of a
crystal but retains the short-range order, where all atoms remain
trapped inside cages formed by their neighbors~\cite{KobAnd95}. When
an amorphous material is strained, the onset of plastic deformation
is governed by the localized collective rearrangements of small
groups of atoms~\cite{Falk98}.   In this process, each rearrangement
creates a long-range elastic field~\cite{Eshelby57} that leads to a
stress redistribution in the system and may drive a nearby region
past its instability threshold, which in turn might trigger
secondary shear transformations~\cite{Lemaitre09}.   In a driven
system, however, this process is difficult to
characterize~\cite{BarratCh8}.     An alternative approach to study
the effects of the local rearrangement of atoms and propagation of
the mechanical signal on the structural relaxation of the material
is to apply an artificial shear transformation in a quiescent
system~\cite{Puosi14}.

\vskip 0.05in

% more on the procedure on strain

In our study, the material was sheared by an artificial inclusion
that consists of about $135$ atoms forming a sphere, which is
located at the center of the simulation cell (see
Fig.\,\ref{fig:snapshot_system}).  The inclusion atoms do not
interact with each other and undergo a reversible displacement
during the time interval $\tau_i$.      In order to maintain the
same density of the material, the inclusion was gradually deformed
from a sphere into an ellipsoid of the same volume and then back to
the sphere with the strain amplitude varied according to
Eq.\,(\ref{Eq:strain}).    We note that the reversible shear
transformation described by Eq.\,(\ref{Eq:strain}) is different from
the shear transformation considered in~\cite{Puosi14}, where a group
of atoms forming an inclusion were instantaneously displaced and
fixed in the new position, thus mimicking an elementary plastic
event occurring in deformed glasses.    In our study, the analysis
of the plastic deformation was based on the averaged atom positions
before and after the shear transformation.    Thus, the control
parameters in the problem include the strain amplitude $\epsilon_0$,
the time scale of the shear transformation $\tau_i$, and the
friction coefficient $\Gamma$ that controls the damping term in
Eq.\,(\ref{Eq:Langevin_x}).

\vskip 0.05in

% plastic elastic and cage detection here

As it was already mentioned in the previous study~\cite{Puosi14}, if
the strain amplitude of the reversible shear transformation of an
artificial inclusion is below a few percent, then the material
deforms elastically and the average relative displacement of atoms
is less than about $0.1\,\sigma$.    In contrast, when the strain
amplitude is sufficiently large, the deformation usually involves
several localized plastic events.     In our study, the strain
amplitude is greater than $0.05$, which typically results in a
finite density of atoms that undergo large displacement upon a
reversible shear transformation.    These large displacements, or
cage jumps, can be identified using the cage detection algorithm
recently introduced by Candelier\,\textit{et al.}~\cite{BiroliPRL09}
and tested for a number of two-dimensional
systems~\cite{BiroliEPL10,BiroliPRL10}.    In this method, the
effective distance between two segments of a particle trajectory is
computed and then compared with a typical cage size.     Similar to
the earlier studies~\cite{Priezjev13,Priezjev14}, in our analysis, a
cage jump was detected if the effective distance between particle
positions before and after the shear transformation is greater than
$0.1\,\sigma$; otherwise, a particle returned to its cage.

\vskip 0.05in

% density profiles of cage jumps and examples

% at small strain amplitudes no cage jumps at all.

\begin{figure}[t]
\includegraphics[width=12.cm,angle=0]{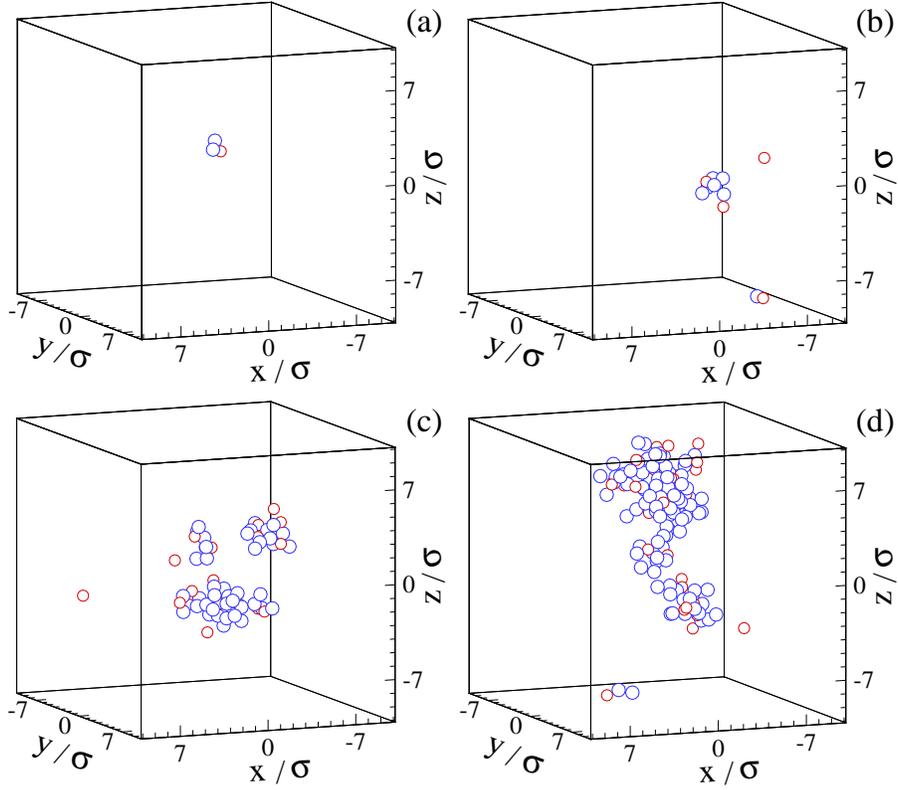}
\caption{(Color online)  Snapshots of cage jump configurations for
the strain amplitudes (a) $\epsilon_0=0.1$, (b) $\epsilon_0=0.2$,
(c) $\epsilon_0=0.3$ and (d) $\epsilon_0=0.4$. The friction
coefficient is $\Gamma=1.0\,\tau^{-1}$ and the time scale of the
shear event is $\tau_i=10\,\tau$.  The artificial inclusion is
located at the center of the simulation cell (not shown).   }
\label{fig:snapshot_clusters}
\end{figure}

Examples of typical cage jump configurations are presented in
Fig.\,\ref{fig:snapshot_clusters} for different strain amplitudes of
the induced shear transformation, which occurred during the time
interval $\tau_i=10\,\tau$ in a regime of intermediate damping
$\Gamma=1.0\,\tau^{-1}$.    It can be seen that with increasing
strain amplitude, the number of cage jumps increases and they tend
to aggregate into compact clusters, although some mobile atoms
appear to be isolated (see Fig.\,\ref{fig:snapshot_clusters}). In
order to quantify the spatial distribution of cage jumps, we
computed the radial density profiles, which were averaged in thin
spherical shells of $0.1\,\sigma$. When the radius of the shell was
larger than half of the simulation box length, the volume of six
spherical caps was subtracted from the volume of the shell to
properly normalize the density profiles.    Due to insufficient
statistics, the data near the corners of the cubic box,
$r>L/\sqrt{2}$, were not reported.

\vskip 0.05in

% density profiles

\begin{figure}[t]
\includegraphics[width=12.cm,angle=0]{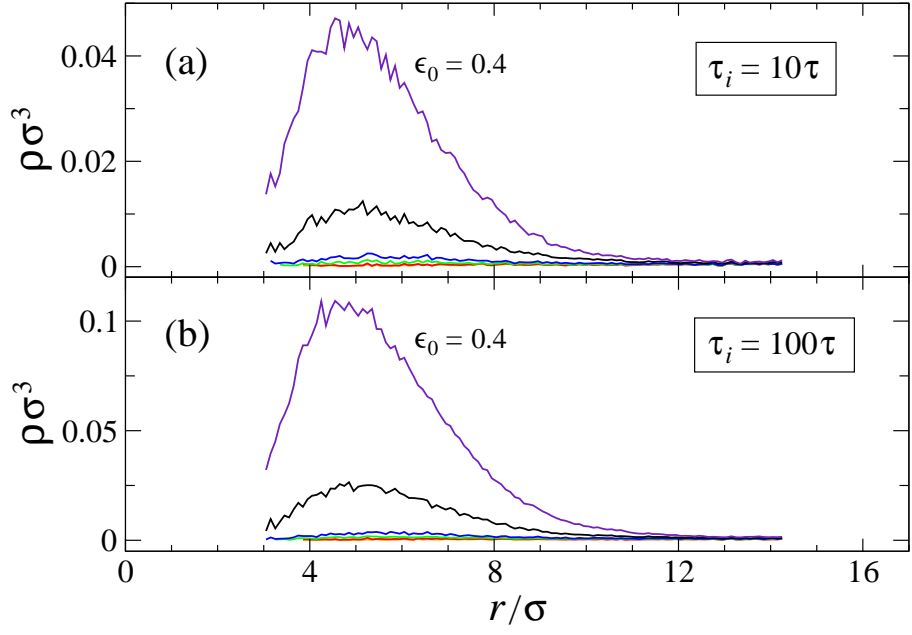}
\caption{(Color online) Averaged density profiles of cage jumps as a
function of the distance from the center of the inclusion for (a)
$\tau_i=10\,\tau$ and (b) $\tau_i=100\,\tau$. The friction
coefficient is $\Gamma=1.0\,\tau^{-1}$ and the strain amplitude is
$\epsilon_0=0.05$, $0.1$, $0.15$, $0.2$, $0.3$, and $0.4$ from
bottom to top. } \label{fig:den_var_eps}
\end{figure}

Figure\,\ref{fig:den_var_eps} shows the representative radial
density profiles of cage jumps for two values of the shear
transformation time scale $\tau_i=10\,\tau$ and $\tau_i=100\,\tau$.
In both cases, the average density of cage jumps becomes larger as
the strain amplitude increases, however, it remains much smaller
than the density of the material, i.e., $\rho=1.2\,\sigma^{-3}$.
Note that, somewhat unexpectedly, the density of cage jumps
increases from $r_i=3\,\sigma$ up to about $5\,\sigma$, despite a
relatively large deformation of the material near the inclusion
during the shear transformation process.     This behavior can be
explained by the fact that the outer surface of the inclusion is
rough on the molecular scale, and, therefore, the atoms of the
material in contact with the inclusion atoms have a part of their
cages to be reversibly deformed, thus reducing the probability of
irreversible displacement of adjacent atoms.    This trend in the
density profiles was observed for all control parameters in our
study.

% finite size effects: elastic deformation travels through the PBC

\vskip 0.05in

% quadrupolar symmetry and power law cluster size

\begin{figure}[t]
\includegraphics[width=12.cm,angle=0]{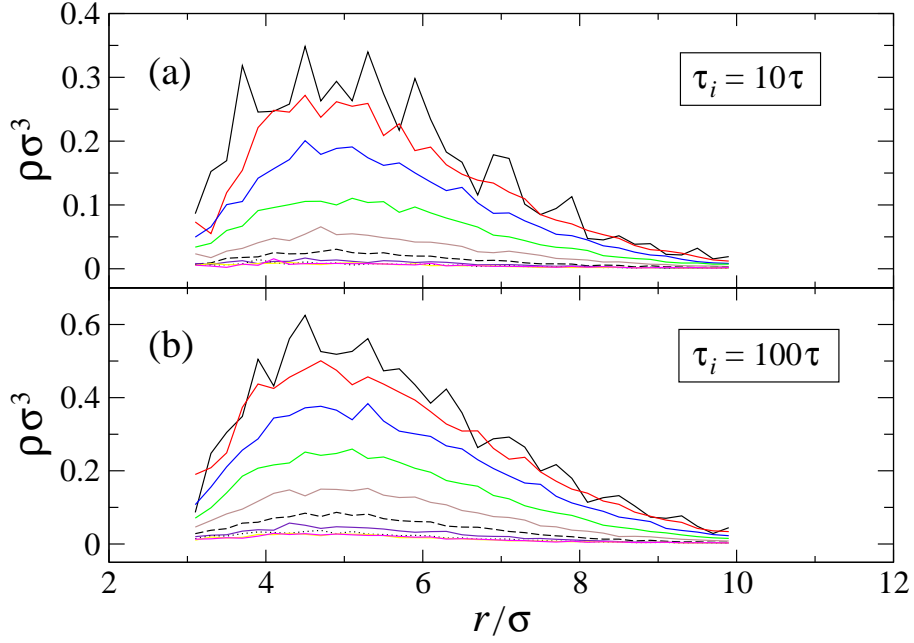}
\caption{(Color online)  Averaged radial density profiles of cage
jumps as a function of the angle $\theta$ with respect to the
$(1,1,1)$ direction (see text for details).  The strain amplitude in
both cases $\epsilon_0=0.4$ and the time scale of the shear
transformation is (a) $\tau_i=10\,\tau$ and (b) $\tau_i=100\,\tau$.
The angle is $\theta=0^{\circ}$, $10^{\circ}$, $20^{\circ}$,
$30^{\circ}$, $40^{\circ}$, $50^{\circ}$, $60^{\circ}$,
$70^{\circ}$, $80^{\circ}$, $90^{\circ}$ from top to bottom. }
\label{fig:den_var_theta}
\end{figure}

\begin{figure}[t]
\includegraphics[width=12.cm,angle=0]{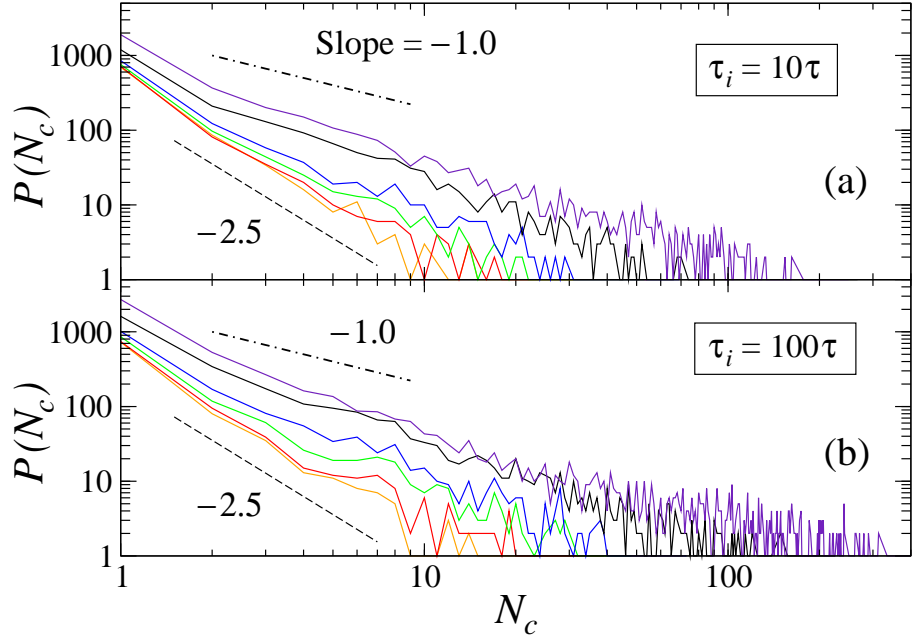}
\caption{(Color online) The probability distribution of cluster
sizes of cage jumps for (a) $\tau_i=10\,\tau$ and (b)
$\tau_i=100\,\tau$.   The strain amplitude is $\epsilon_0=0.05$,
$0.1$, $0.15$, $0.2$, $0.3$, and $0.4$ from bottom to top. The
friction coefficient is $\Gamma=1.0\,\tau^{-1}$ in both cases. The
straight lines with the slope of $-2.5$ (dashed lines) and $-1$
(dash-dotted lines) are plotted for reference.  }
\label{fig:cluster_size_eps}
\end{figure}

During the shear transformation process, the displacement field
around the inclusion has a quadrupolar symmetry with respect to the
major axis of the ellipsoid~\cite{Procaccia13}, and, therefore, the
distribution of cage jumps is not expected to be spatially
isotropic.      We next computed the radial density of cage jumps
but averaged it within hollow cones with apex angles ranging from
$2\,\theta-10^{\circ}$ to $2\,\theta+10^{\circ}$.
Figure\,\ref{fig:den_var_theta} shows the radial density profiles of
cage jumps for different values of the angle $\theta$ measured with
respect to the major axis of the ellipsoid, which is oriented
parallel to the vector $(1, 1, 1)$.   It could be seen that the data
are somewhat noisy despite a relatively large strain amplitude
$\epsilon_0=0.4$.     However, it is clear that the orientational
dependence of the density profiles is highly anisotropic; the
density is larger (smaller) in the direction parallel
(perpendicular) to the major axis of the ellipsoid.    Furthermore,
as illustrated in Fig.\,\ref{fig:snapshot_clusters}, cage jumps tend
to aggregate into clusters.   The distribution of cluster sizes is
reported in Fig.\,\ref{fig:cluster_size_eps} for $\tau_i=10\,\tau$
and $\tau_i=100\,\tau$.     It is observed that cluster sizes are
roughly power-law distributed with the slope varying from $-2.5$ at
small strain amplitudes to about $-1.0$ for $\epsilon_0\gtrsim 0.3$
with the largest clusters of several hundred atoms.

\vskip 0.05in

% time scale of shear transformation

\begin{figure}[t]
\includegraphics[width=12.cm,angle=0]{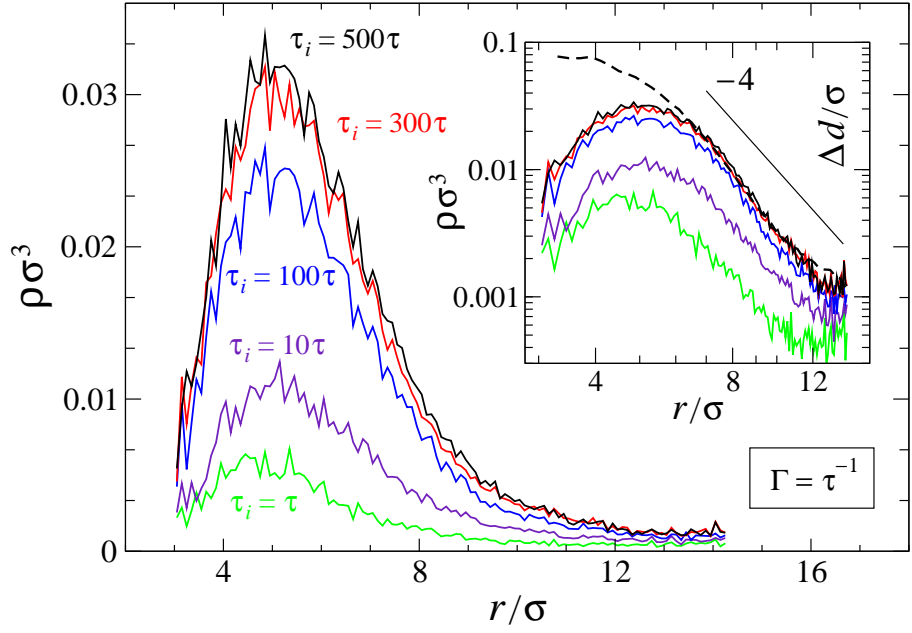}
\caption{(Color online)  Averaged density profiles of cage jumps as
a function of the distance from the center of the inclusion for the
strain amplitude $\epsilon_0=0.3$ and the friction coefficient
$\Gamma=1.0\,\tau^{-1}$.    The time scale of the shear event is
$\tau_i/\tau=1$, $10$, $100$, $300$, and $500$.  Inset: the same
density profiles are plotted on a logarithmic scale. The local
deformation $\Delta d$ is denoted by the dashed curve (see text for
details).  The scale for $\Delta d/\sigma$ is the same as for
$\rho\sigma^3$.   The straight line with a slope $-4$ is shown for
reference. } \label{fig:den_var_timescale}
\end{figure}

The local deformation of the material that facilitates cage jumps
depends on how the mechanical signal propagates in the system and
the time scale of the induced shear transformation.   In
Figure\,\ref{fig:den_var_timescale}, we plot the radial density
profiles of cage jumps for different values of $\tau_i$ in the
regime of intermediate damping $\Gamma=1.0\,\tau^{-1}$.    It is
evident that the density of cage jumps increases with increasing
shear transformation time scale.     When $\tau_i$ is small, then
the amplitude of the local displacement field in the material during
the shear event is smaller than it would be in the limit of
quasistatic deformation; and, therefore, the probability of
cage-to-cage jumps is reduced.    In other words, for a given value
of the strain amplitude $\epsilon_0$, the largest particle
displacement in the material occurs when the time scale of the shear
transformation is larger than the time scale of the propagation of
the deformation in the system. As shown in
Fig.\,\ref{fig:den_var_timescale}, with further increasing time
scale of the shear event, i.e., $\tau_i \gtrsim 300\,\tau$, the
density of cage jumps remains unchanged (within statistical
uncertainty).

\vskip 0.05in

% finite-amplitude shear transformation

In order to establish a correlation between the density of cage
jumps and deformation of the material during the shear event, we
performed a set of separate simulations where the shear strain given
by Eq.\,(\ref{Eq:strain}) was varied during the time interval
$0\leqslant t \leqslant 0.5\,\tau_i$ up to a maximum value of
$\epsilon_0$.    The numerical procedure was applied as follows.
Similar to the case of the reversible shear transformation described
in Sec.\,\ref{sec:MD_Model}, the binary glass was first equilibrated
while the inclusion atoms were fixed within a sphere of radius
$r_i=3\,\sigma$.    Next, the sphere was transformed into an
ellipsoid of the same volume according to Eq.\,(\ref{Eq:strain})
during the time interval $0\leqslant t \leqslant 0.5\,\tau_i =
100\,\tau$. When the maximum strain $\epsilon_0$ was attained, the
inclusion atoms were fixed again and the system was allowed to
evolve for $10^3\,\tau$.     Then, the average atom positions were
collected and the relative displacement of atoms during the
sphere-to-ellipsoid transformation was analyzed.

\vskip 0.05in

The local deformation of the material after a sphere-to-ellipsoid
transformation deviates from the elastic behavior at sufficiently
large values of $\epsilon_0$ and typically involves several plastic
events.    Here, we introduce a measure of the local deformation,
$\Delta d$, based on the relative displacement of neighboring
particles.    In the undeformed glass, we first identify all
tetrahedra formed by nearest-neighbor atoms with mutual distance
less than $r_d=1.2\,\sigma$.     For each tetrahedron, the
difference between the longest and shortest edges, $d$, is computed.
After the sphere-to-ellipsoid transformation, the quantity $d$ is
determined again for the same tetrahedra.     Finally, the measure
of the local deformation, $\Delta d$, is calculated for each
tetrahedron by subtracting $d$ in the undeformed state from $d$
after the transformation, and then averaged over all tetrahedra and
all realizations of disorder.     In this definition, the quantity
$\Delta d$ is insensitive to a pure rotation or translation of
tetrahedra but it takes into account the relative displacement of
neighboring atoms.

\vskip 0.05in

% on a log-log scale

The radial dependence of the local deformation, $\Delta d\,(r)$, is
plotted on a log-log scale in the inset of
Fig.\,\ref{fig:den_var_timescale} along with the density profiles of
cage jumps for $\epsilon_0=0.3$.    It can be seen that the decay of
the density profiles for different values of $\tau_i$ correlates
well with $\Delta d\,(r)$ for $r \gtrsim 6\,\sigma$. These results
suggest that formation of cage jumps during the reversible shear
event is determined by the amplitude of the local deformation of the
material.    Note that the density profiles are nearly constant at
large distances $r \gtrsim L/2$ due to periodic boundary conditions.
We also comment that the slope of $\Delta d\,(r)$ is rather
insensitive to the exact value of $r_d$ in the range $1.1\,\sigma
\leqslant r_d \leqslant 1.5\,\sigma$.

\vskip 0.05in

% overdamped versus underdamped dynamics

\begin{figure}[t]
\includegraphics[width=12.cm,angle=0]{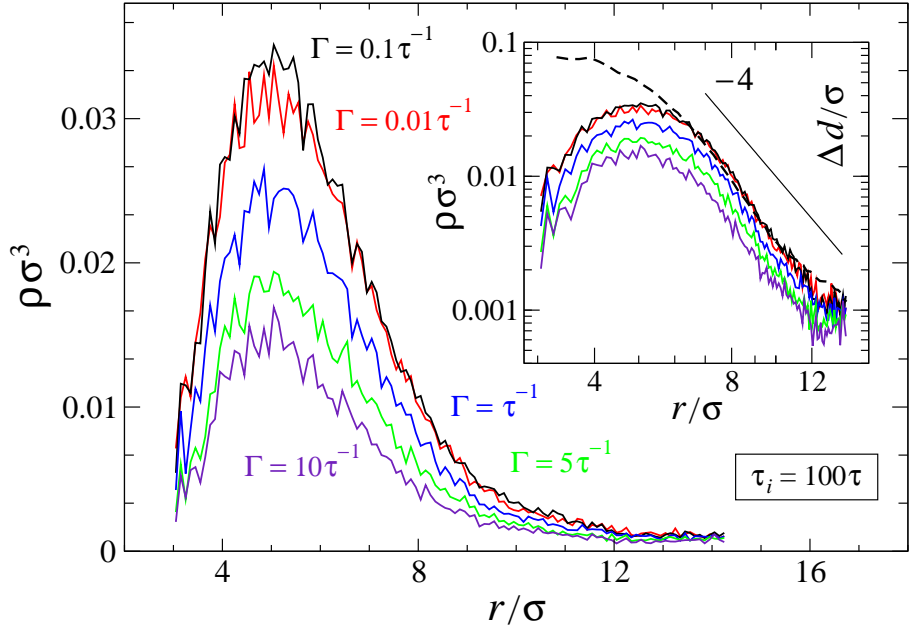}
\caption{(Color online)  Averaged density profiles of cage jumps as
a function of the distance from the center of the cell for
$\epsilon_0=0.3$ and $\tau_i=100\,\tau$.   The friction coefficient
is $\Gamma\tau=0.01$, $0.1$, $1$, $5$, $10$.   The inset shows the
same density profiles on a log-log scale.  The radial dependence of
$\Delta d$ is the same as in the inset of
Fig.\,\ref{fig:den_var_timescale} (dashed curve).  The scale for
$\Delta d/\sigma$ and $\rho\sigma^3$ is the same.   The black line
with a slope $-4$ is plotted as a reference.  }
\label{fig:den_var_friction}
\end{figure}

Furthermore, the effect of inertia on the density profiles of cage
jumps is shown in Fig.\,\ref{fig:den_var_friction} for the
relatively large strain amplitude $\epsilon_0=0.3$ and time scale
$\tau_i=100\,\tau$ in order to obtain better statistics.    It is
observed that with decreasing friction coefficient, the density of
cage jumps increases and it appears to saturate for $\Gamma \lesssim
0.1\,\tau^{-1}$.       In the strongly damped cases,
$\Gamma=5\,\tau^{-1}$ and $10\,\tau^{-1}$, the propagation of the
displacement field occurs at times longer than $\tau_i$, which
results in a smaller amplitude of the local strain and thus lower
density of cage jumps.      Similar transient behavior of the
elastic response to a local shear transformation was reported in a
two-dimensional model glass in the overdamped regime~\cite{Puosi14}.
In the case of weakly damped dynamics, $\Gamma\lesssim
0.1\,\tau^{-1}$,  the transmission of the elastic signal in the
system is faster than the duration of the shear event, which leads
to the largest amplitude of the displacement field and highest
density of cage jumps (shown in Fig.\,\ref{fig:den_var_friction}).

\vskip 0.05in

% The data can be collapsed onto a master curve by rescaling the vertical axis (not shown).

\begin{figure}[t]
\includegraphics[width=12.cm,angle=0]{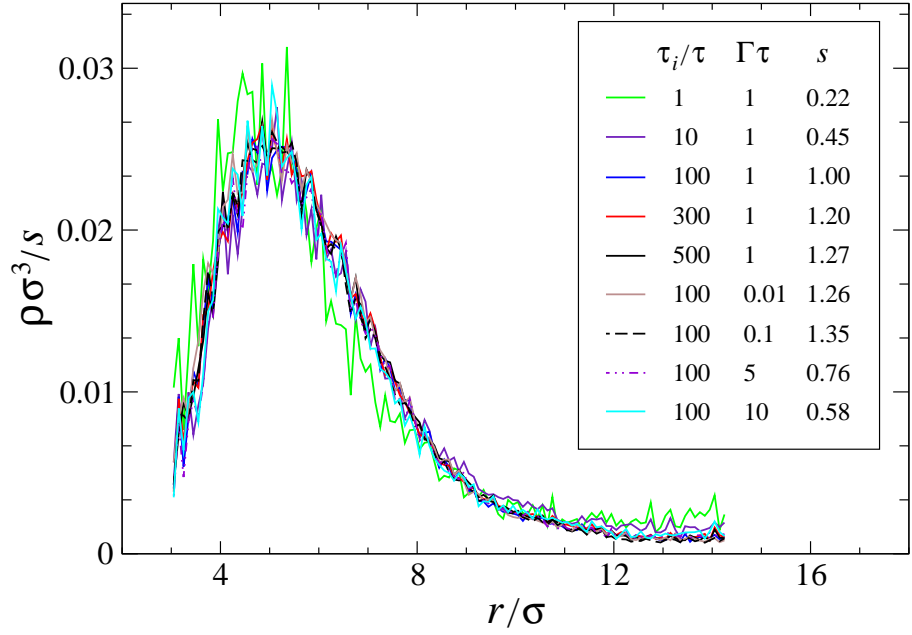}
\caption{(Color online) Collapse of the density profiles of cage
jumps for the strain amplitude $\epsilon_0=0.3$.  The data are the
same as in Figs.\,\ref{fig:den_var_timescale} and
\ref{fig:den_var_friction}.  The values of the friction coefficient
$\Gamma$, the time scale of the shear transformation $\tau_i$, and
the scaling factor $s$ are tabulated in the inset.   }
\label{fig:density_collapse}
\end{figure}

Finally, the data reported in Figs.\,\ref{fig:den_var_timescale} and
\ref{fig:den_var_friction} can be made to collapse onto a single
master curve by plotting the density profiles as $\rho(r)/s$, where
$s$ is a scaling factor.     The rescaled density profiles are shown
in Fig.\,\ref{fig:density_collapse} for the indicated values of
$\Gamma$, $\tau_i$ and $s$.     The collapse of the data is quite
good, except for the case $\tau_i=\tau$ and $\Gamma=\tau^{-1}$,
where the amplitude of the density profile is relatively small.  The
results in Fig.\,\ref{fig:density_collapse} suggest that the density
profiles as a function of the distance $r$, the friction coefficient
$\Gamma$, and the time scale of the shear event $\tau_i$ obey the
remarkable factorization
\begin{equation}
\rho(r,\Gamma,\tau_i)= s(\Gamma,\tau_i)\,f(r)
\label{Eq:scaling}
\end{equation}
in a wide range of parameters $\Gamma$ and $\tau_i$.   Thus, the
shape of the density profiles is determined by the universal
function $f(r)$, which exhibits a maximum within about two molecular
diameters from the surface of the inclusion and follows a power-law
decay at larger distances.    Interestingly, the variation of the
scaling factor $s$ is relatively small despite the fact that
parameters $\Gamma$ and $\tau_i$ vary over about three orders of
magnitude.

\section{Conclusions}

In summary, molecular dynamics simulations were performed to
investigate the effect of a local shear transformation on plastic
deformation of an amorphous solid.   We considered a spherical
inclusion, which was transformed into an ellipsoid with the major
axis along one of the diagonals of a cubic box and then converted
back into the sphere. At sufficiently large strain amplitudes, the
deformation of the material typically involved several plastic
events, which were analyzed based on the averaged atom positions
before and after the shear transformation.   Using the cage
detection algorithm, the large particle displacements, or cage
jumps, were identified and their spatial distribution was studied as
a function of the strain amplitude, the time scale of the shear
transformation, and the friction coefficient that controls the
damping in the system.

\vskip 0.05in

We found that, in general, the density of irreversible cage jumps
increases with increasing strain amplitude of the shear
transformation.    In a narrow region within about two particle
diameters from the surface of the inclusion, the probability of cage
jumps is reduced because of the strong influence of the inclusion
atoms that undergo reversible displacements.     It was also shown
that cage jumps tend to aggregate into clusters that are
approximately power-law distributed.     The density profiles of
cage jumps decay away from the center of the inclusion, which agrees
well with the radial dependence of the local deformation of the
material.      Furthermore, due to the quadrupolar symmetry of the
displacement field, the distribution of cage jumps is highly
anisotropic with a preferred direction along the major axis of the
ellipsoid.      For a given strain amplitude, the density of cage
jumps increases upon either increasing time scale of the shear event
or decreasing friction coefficient.     These trends were
rationalized in terms of the maximum amplitude of the displacement
field in the material during the shear transformation.

\section*{Acknowledgments}

Financial support from the National Science Foundation
(CBET-1033662) is gratefully acknowledged.  Computational work in
support of this research was performed at Michigan State
University's High Performance Computing Facility and the Ohio
Supercomputer Center.

%%%%%%%%%%%%%%% FIGURES %%%%%%%%%%%%%%%%%%%%%%%

% snapshot Present_pdf

% typical clusters of jumps

% density profiles varying strain amplitude

% density profiles for different angles theta eps=0.3 # 13 and 20

% cluster size distribution

% density profiles varying timescale

% density profiles varying friction coefficient

% density profiles collapse

\bibliographystyle{prsty}

\begin{thebibliography}{99}

% review
\bibitem{FalkRev10}   M.~L. Falk and C.~E. Maloney,
                      % Simulating the mechanical response of amorphous solids using atomistic methods
                      Eur. Phys. J. B {\bf 75}, 405 (2010).
% review
\bibitem{ChengMa11}   Y.~Q. Cheng and E. Ma,
                      % Atomic-level structure and structure–property relationship in metallic glasses
                      Prog. Mater Sci. {\bf 56}, 379 (2011).

\bibitem{BarratCh8}   J.-L. Barrat and A. Lemaitre,
                      ``Heterogeneities in amorphous systems under shear," Chapter 8 in the book
                      \textit{Dynamical Heterogeneities in Glasses, Colloids, and Granular Media}
                      (Oxford University Press, 2011).

\bibitem{Argon79}     A.~S. Argon, Acta Metall. {\bf 27}, 47 (1979).
                      % Plastic deformation in metallic glasses

\bibitem{ArgonKuo}    A. Argon and H. Kuo, Mater. Sci. Eng. {\bf 39}, 101 (1979).
                      % Plastic flow in a disordered bubble raft (an analog of a metallic glass)

% STZ
\bibitem{Falk98}      M.~L. Falk and J.~S. Langer, Phys. Rev. E {\bf 57}, 7192 (1998).
                      % Dynamics of viscoplastic deformation in amorphous solids

% local density
\bibitem{Bailey06}    N.~P. Bailey, J. Schiotz, and K.~W. Jacobsen,
                      % Atomistic simulation study of the shear-band deformation mechanism in Mg-Cu metallic glasses
                      Phys. Rev. B {\bf 73}, 064108 (2006).

% local modulus
\bibitem{dePablo04}   K. Yoshimoto, T.~S. Jain, K. van Workum, P.~F. Nealey, and J.~J. de Pablo,
                      % Mechanical Heterogeneities in Model Polymer Glasses at Small Length Scales
                      Phys. Rev. Lett. {\bf 93}, 175501 (2004).

% local modulus
\bibitem{Tsamados09}  M. Tsamados, A. Tanguy, C. Goldenberg, and J.-L. Barrat,
                      % Local elasticity map and plasticity in a model Lennard-Jones glass
                      Phys. Rev. E {\bf 80}, 026112 (2009).

% local modulus
\bibitem{Procaccia10} S. Karmakar, A. Lemaitre, E. Lerner, and I. Procaccia,
                      % Predicting Plastic Flow Events in Athermal Shear-Strained Amorphous Solids
                      Phys. Rev. Lett. {\bf 104}, 215502 (2010).

% short range order
\bibitem{Teichler02}  K. Brinkmann and H. Teichler,
                      % Flow state in molecular-dynamics-simulated deformed amorphous Ni0.5Zr0.5
                      Phys. Rev. B {\bf 66}, 184205 (2002).

% short range order
\bibitem{Albano05}    F. Albano and M. Falk, J. Chem. Phys. {\bf 122}, 154508 (2005).
                      % Shear softening and structure in a simulated three-dimensional binary glass

% vib
\bibitem{Tanguy10}    A. Tanguy, B. Mantisi, and M. Tsamados,
                      % Vibrational modes as a predictor for plasticity in a model glass
                      Europhys. Lett. {\bf 90}, 16004 (2010).
% vib
\bibitem{Manning11}   M.~L. Manning and A.~J. Liu,
                      % Vibrational Modes Identify Soft Spots in a Sheared Disordered Packing,
                      Phys. Rev. Lett. {\bf 107}, 108302 (2011).
% vib
\bibitem{Schoenholz14} S.~S. Schoenholz, A.~J. Liu, R.~A. Riggleman, and J. Rottler,
                       % Understanding Plastic Deformation in Thermal Glasses from Single-Soft-Spot Dynamics
                       Phys. Rev. X {\bf 4}, 031014 (2014).


% AQS
\bibitem{Maloney06}   C.~E. Maloney and A. Lemaitre,
                      % Amorphous systems in athermal, quasistatic shear
                      Phys. Rev. E {\bf 74}, 016118 (2006).
% AQS
\bibitem{Tanguy06}    A. Tanguy, F. Leonforte, and J.-L Barrat,
                      Eur. Phys. J. E {\bf 20}, 355 (2006).
                      % Plastic response of a 2D Lennard-Jones amorphous solid:
                      % Detailed analysis of the local rearrangements at very slow strain rate
% AQS
\bibitem{Tsamados10}  M. Tsamados,
                      % Plasticity and dynamical heterogeneity in driven glassy materials
                      Eur. Phys. J. E {\bf 32}, 165 (2010).

% effect inertia
\bibitem{Maloney12}  K.~M. Salerno, C.~E. Maloney, and M.~O. Robbins,
                     % Avalanches in Strained Amorphous Solids: Does Inertia Destroy Critical Behavior?
                     Phys. Rev. Lett. {\bf 109}, 105703 (2012).

% model for correlated flips
\bibitem{Bulatov94}   V.~V. Bulatov and A.~S. Argon,
                      Model. Simul. Mater. Sci. Eng. {\bf 2}, 167 (1994).
                      % A stochastic model for continuum elasto-plastic behavior: I.
                      % Numerical approach and strain localization

\bibitem{Eshelby57}   J.~D. Eshelby,
                      % The determination of the elastic field of an ellipsoidal inclusion, and related problems
                      Proc. Roy. Soc. London A {\bf 241}, 376 (1957).

\bibitem{Lemaitre09}  A. Lemaitre and C. Caroli,
                      % Rate-Dependent Avalanche Size in Athermally Sheared Amorphous Solids
                      Phys. Rev. Lett. {\bf 103}, 065501 (2009).

% 2d inclusion
\bibitem{Puosi14}     F. Puosi, J. Rottler, and J.-L. Barrat,
                      % Time dependent elastic response to a local shear transformation in amorphous solids
                      Phys. Rev. E {\bf 89}, 042302 (2014).

% cyclic
\bibitem{Arratia13}    N.~C. Keim and P.~E. Arratia,
                       % Yielding and microstructure in a 2D jammed material under shear deformation
                       Soft Matter {\bf 9}, 6222 (2013).

% cyclic
\bibitem{Arratia14}    N.~C. Keim and P.~E. Arratia,
                       % Mechanical and Microscopic Properties of the Reversible Plastic Regime
                       % in a 2D Jammed Material
                       Phys. Rev. Lett. {\bf 112}, 028302 (2014).
% cyclic
\bibitem{Spaepen14}    K.~E. Jensen, D.~A. Weitz, and F. Spaepen,
                       % Local shear transformations in deformed and quiescent hard-sphere colloidal glasses
                       Phys. Rev. E {\bf 90}, 042305 (2014).

% cyclic
\bibitem{Cipelletti14} E.~D. Knowlton, D.~J. Pine, and L. Cipelletti,
                       % A microscopic view of the yielding transition in concentrated emulsions
                       Soft Matter {\bf 10}, 6931 (2014).

% MD
\bibitem{Sastry13}    D. Fiocco, G. Foffi, and S. Sastry,
                      % Oscillatory athermal quasistatic deformation of a model glass
                      Phys. Rev. E {\bf 88}, 020301(R) (2013).

% MD
\bibitem{Priezjev13}  N.~V. Priezjev,
                      % Heterogeneous relaxation dynamics in amorphous materials under cyclic loading
                      Phys. Rev. E {\bf 87}, 052302 (2013).
% MD
\bibitem{Reichhardt13} I. Regev, T. Lookman, and C. Reichhardt,
                       % Onset of irreversibility and chaos in amorphous solids under periodic shear
                       Phys. Rev. E {\bf 88}, 062401 (2013).
% MD
\bibitem{Priezjev14}  N.~V. Priezjev,
                      % Dynamical heterogeneity in periodically deformed polymer glasses
                      Phys. Rev. E {\bf 89}, 012601 (2014).

% MD DETAILS

\bibitem{KobAnd95}    W. Kob and H.~C. Andersen,
                      %  TESTING MODE-COUPLING THEORY FOR A SUPERCOOLED BINARY LENNARD-JONES MIXTURE -
                      %  THE VAN HOVE CORRELATION-FUNCTION
                      Phys. Rev. E {\bf 51}, 4626 (1995).

\bibitem{Allen87}     M.~P. Allen and D.~J. Tildesley,
                      {\it Computer Simulation of Liquids} (Clarendon, Oxford, 1987).

% Cage Algorithm
\bibitem{BiroliPRL09} R. Candelier, O. Dauchot, and G. Biroli,
                      % Building Blocks of Dynamical Heterogeneities in Dense Granular Media
                      Phys. Rev. Lett. {\bf 102}, 088001 (2009).

% EXP dense glassy granular media fluidized bed     Cage Algorithm
\bibitem{BiroliEPL10} R. Candelier, O. Dauchot, and G. Biroli,
%                     % Dynamical facilitation decreases when approaching the granular glass transition
                      Europhys. Lett. {\bf 92}, 24003 (2010).

% MD two-dimensional glass-forming liquid    Cage Algorithm
\bibitem{BiroliPRL10} R. Candelier, A. Widmer-Cooper, J.~K. Kummerfeld,
                      O. Dauchot, G. Biroli, P. Harrowell, and D.~R. Reichman,
                      % Spatiotemporal Hierarchy of Relaxation Events, Dynamical Heterogeneities,
                      % and Structural Reorganization in a Supercooled Liquid
                      Phys. Rev. Lett. {\bf 105}, 135702 (2010).

% elastic fields in 3d
\bibitem{Procaccia13} R. Dasgupta, O. Gendelman, P. Mishra, I. Procaccia, and C.~A.~B.~Z. Shor,
                      % Shear localization in three-dimensional amorphous solids
                      Phys. Rev. E {\bf 88}, 032401 (2013).

\end{thebibliography}

\end{document}